\documentclass[10pt,letterpaper]{article}
\usepackage{opex3}
\usepackage{hyperref}

\begin{document}

\title{Enhanced frequency up-conversion in Rb vapor}

\author{A.\ Vernier,$^1$ S.\ Franke-Arnold,$^1$ E.\ Riis,$^2$ and A.\ S.\ Arnold$^2$}

 \address{$^1$SUPA, Dept.\ of Physics and Astronomy, University of Glasgow, Glasgow G12 8QQ, UK\\
 $^2$SUPA, Dept.\ of Physics, University of Strathclyde, Glasgow G4 0NG, UK}

\begin{abstract}
We demonstrate highly efficient generation of coherent $420\,$nm light via up-conversion of near-infrared lasers in a hot rubidium vapor cell. By
optimizing pump polarizations and frequencies we achieve a single-pass conversion efficiency of $260\%$ per Watt, significantly higher than in
previous experiments.  A full exploration of the coherent light generation and fluorescence as a function of both pump frequencies reveals that
coherent blue light is generated close to $^{85}$Rb two-photon resonances, as predicted by theory, but at high vapor pressure is suppressed in
spectral regions that do not support phase matching or exhibit single-photon Kerr refraction. Favorable scaling of our current $1\,$mW blue beam
power with additional pump power is predicted.
\end{abstract}


Nonlinear optical processes can be greatly enhanced for quasi-resonant atomic and molecular systems, allowing phenomena like efficient frequency
up-conversion, four-wave mixing, slow light or image storage to be studied at low light intensities. On resonance, absorption hinders the build-up of
strong coherences; this can be counteracted by excitation via two-photon resonances and in particular by using electromagnetically induced
transparency (EIT). Here we report the generation of $1.1\,$mW of $420\,$nm blue light by enhanced frequency up-conversion in a hot rubidium vapor.
This is made possible by long lived two-photon coherences at frequencies that allow propagation under phase-matching conditions. Depending on the
polarization of the pump lasers, up-conversion can be enhanced or almost completely suppressed.

\begin{figure}[!b]\begin{center}
 \mbox{\includegraphics[clip,width=.626\columnwidth]{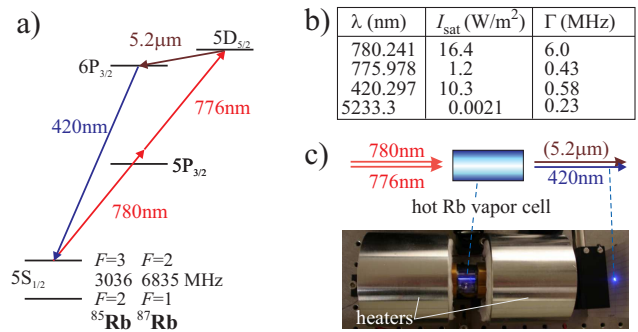}}\end{center}
 \caption{a) The Rb energy level scheme. b) Relevant Rb transition parameters \cite{RbRef}. c) Experimental image showing how co-propagating focussed 780 and $776\,$nm laser beams create a coherent $420\,$nm (and invisible $5.2\,\mu$m) beam. \label{levels}}
 \end{figure}

Alkali metal vapors are versatile tools for spectroscopy and nonlinear optics in atomic physics and have long been used for studies of 2-photon
spectroscopy \cite{2phot,2phot2,2phot3}, dispersion \cite{GeaBanEIT} and EIT \cite{GeaBanEIT,IfanEIT}. In both isotopes of rubidium, the
5S$_{1/2}-5$P$_{3/2}$ (780$\,$nm) and 5P$_{3/2}-5$D$_{5/2}$ (776$\,$nm) transitions (Fig.\ \ref{levels}a) are easily accessible with simple diode
laser systems \cite{arnold}. The extremely strong dipole moment of the infrared 6P$_{3/2}-5$D$_{5/2}$ transition (Fig.\ \ref{levels}b) means that
two-photon pumping with 776$\,$ and 780$\,$nm light facilitates three-photon coherence between ground state and the 6P$_{3/2}$ level. Excitation and
decay via other levels (5P$_{1/2}$, 6P$_{1/2},$ 5D$_{3/2}$) is excluded by large detuning and selection rules. Similar alkali metal transitions have
proven ideal for studies of amplified spontaneous emission versus four-wave mixing \cite{boyd, hamadani}, superfluorescence \cite{blueSF},
multi-photon ionization processes \cite{hamadani}, and sensitive atomic imaging \cite{imagScholten,imagFreegarde}.

Recent papers \cite{hollberg,scholten,hannaford} have reported the generation of tens of $\mu$Watts of coherent 420$\,$nm blue light by pumping the
Rb 5S$-$5P$-$5D transition. In our experiment we obtained up to $1.1\,$mW of coherent blue light for similar input pump powers.  We have studied and
optimized blue light generation for a variety of experimental parameters, namely input beam polarization and frequency, and Rb vapor pressure. In our
system we have measured a maximum of $1.1\,$mW of coherent blue light close to two-photon resonance of the input lasers at
$\Delta_{780}=-\Delta_{776}\cong 1.6\,$GHz. Optimal conditions include a Rb vapor pressure of $10^{-3}\,$mbar and co-circularly polarized input beams
at maximum available powers of $17\,$mW at $776\,$nm and $25\,$mW at $780\,$nm. This corresponds to a conversion efficiency of $\eta=P_{420}/(P_{776}
P_{780})\sim 260\%/$W.

\begin{figure}[!t]
 \begin{minipage}{.53\columnwidth}\mbox{\includegraphics[clip,width=.99\columnwidth]{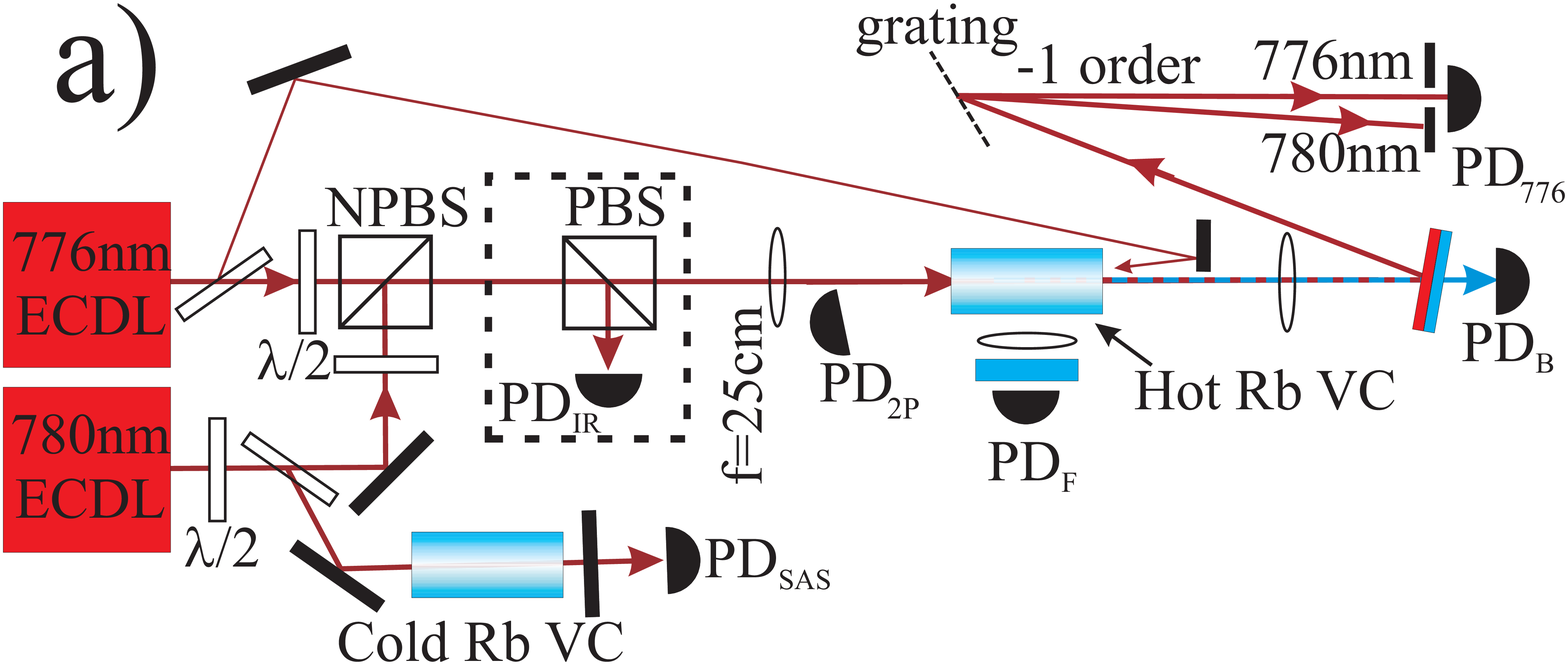}}\end{minipage}
 \begin{minipage}{.055\columnwidth}\end{minipage}
 \begin{minipage}{.41\columnwidth}\mbox{\includegraphics[clip,width=.99\columnwidth]{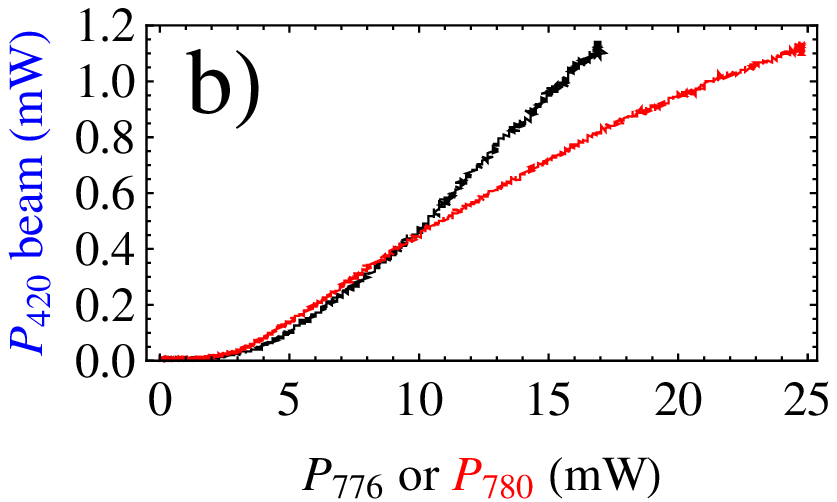}}\end{minipage}
 \caption{a) Experimental schematic, abbreviations used are: PD (photodiode), VC (vapor cell) N/PBS (non/polarizing beamsplitter) and ECDL
(external cavity diode laser). b)~Power in the coherent $420\,$nm beam as a function of $780\,$nm (red) and $776\,$nm (black) input power. For each
trace the power of the other input beam was maximal and detunings and polarization were optimized, see text.\label{setup}}
 \end{figure}

The essence of the experimental setup is shown in Fig.\ \ref{levels}c), with a more detailed view in Fig.\ \ref{setup}a). Laser beams from
Faraday-isolated $776\,$nm and $780\,$nm extended-cavity diode lasers are overlapped on a non-polarizing beamsplitter and focussed into a $75\,$mm
long Rb vapor cell. The pump beams have maximum powers of $17\,$mW and $25\,$mW respectively which for elliptical beams with waists of $0.6\,$mm
$\times$ $1.1\,$mm yield an estimated focal intensity of $\approx 2\times10^6\,$W/m$^2$ each. Note that the $780\,$nm beam area after the cell varies
with frequency, differing by more than an order of magnitude due to Kerr lensing.

We have measured the conversion efficiency as a function of the $776\,$ and $780\,$nm input power (Fig.~\ref{setup}b), with all other parameters
fixed at their optimum values stated above. We note that conversion efficiency depends on the Rabi frequencies and detunings of the pump beams, so
that for low pump powers the chosen detunings are not ideal. The pump power was varied by rotating half wave plates in the individual input beams,
and monitored via the polarizing beamsplitter (PBS) shown in the dashed region of Fig.\ \ref{setup}a). Although the blue light begins to saturate
with $780\,$nm input power, it is linear in $776\,$nm input power in the regions accessible by our diode lasers, promising a further increase in blue
light power for stronger $776\,$nm pumping.

The atomic density and associated vapor temperature were measured to high resolution via absorption spectroscopy of a low power ($5\,\mu$W) $780\,$nm
probe beam \cite{ifan}. This avoided any inaccuracy due to the inhomogeneity of the external cell temperature. For our setup optimal blue light
generation occurs at a temperature of $120\pm 1^{\circ}$C, corresponding to a Rb vapor pressure of $9\times10^{-4}\,$mbar, nearly 4 orders of
magnitude higher than at room temperature. While higher temperatures, and therefore atomic densities, facilitate the nonlinear up-conversion process,
they also increase absorption of the pump and output beams. Overall, the system is fairly robust: more than half of the maximum blue beam efficiency
could be achieved over a temperature range of $104-131^{\circ}$C corresponding to a factor of 5 in Rb pressure, and focussing the near-infrared light
from 0.4 to 4 times the optimal free-space focal intensity yielded similar ($>75\%$ of optimal) conversion efficiency.

\begin{figure}[!b]
\begin{minipage}{.314\columnwidth} \mbox{\includegraphics[clip,width=\columnwidth]{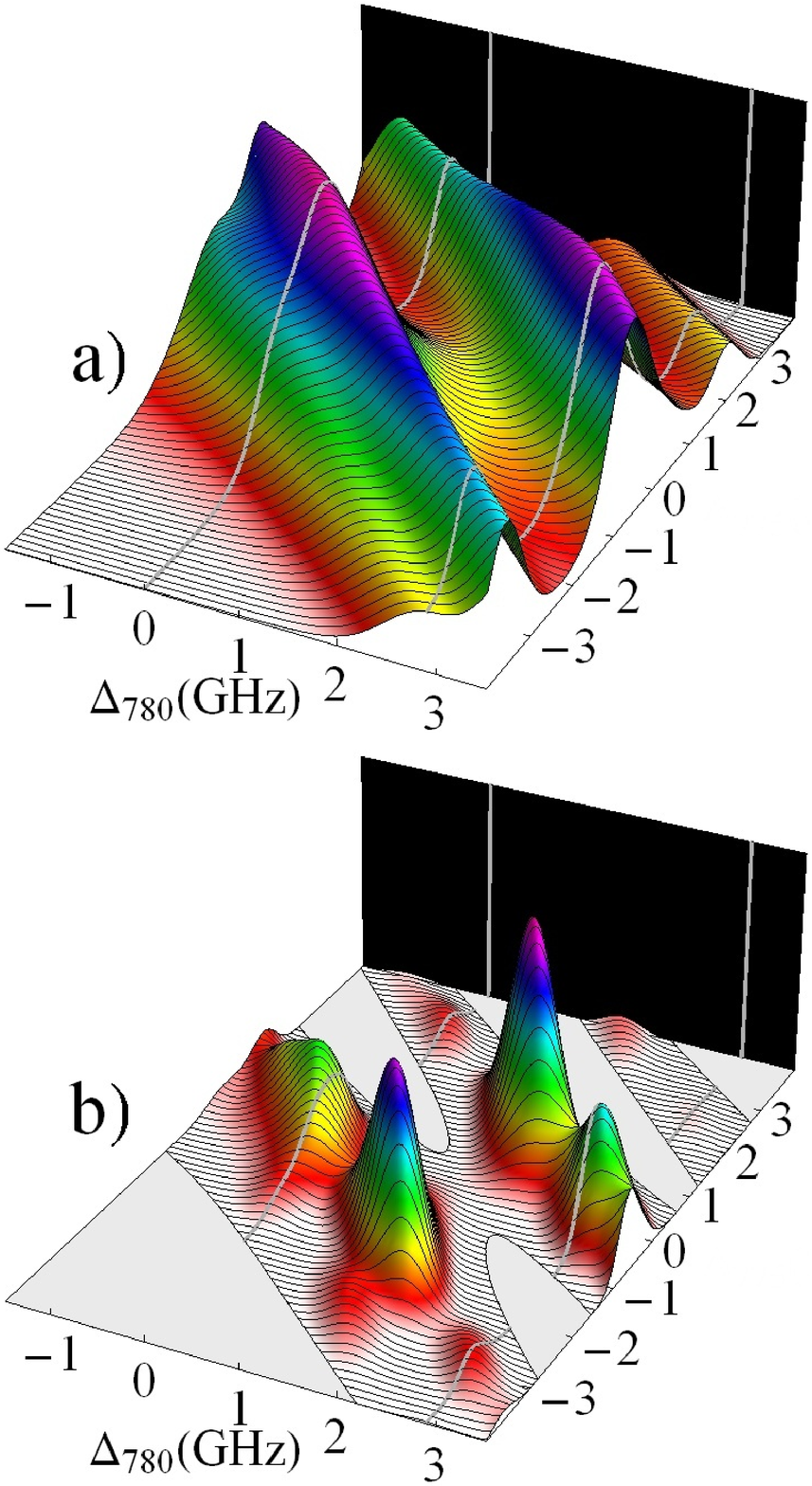}} \end{minipage}
\begin{minipage}{.314\columnwidth} \mbox{\includegraphics[clip,width=\columnwidth]{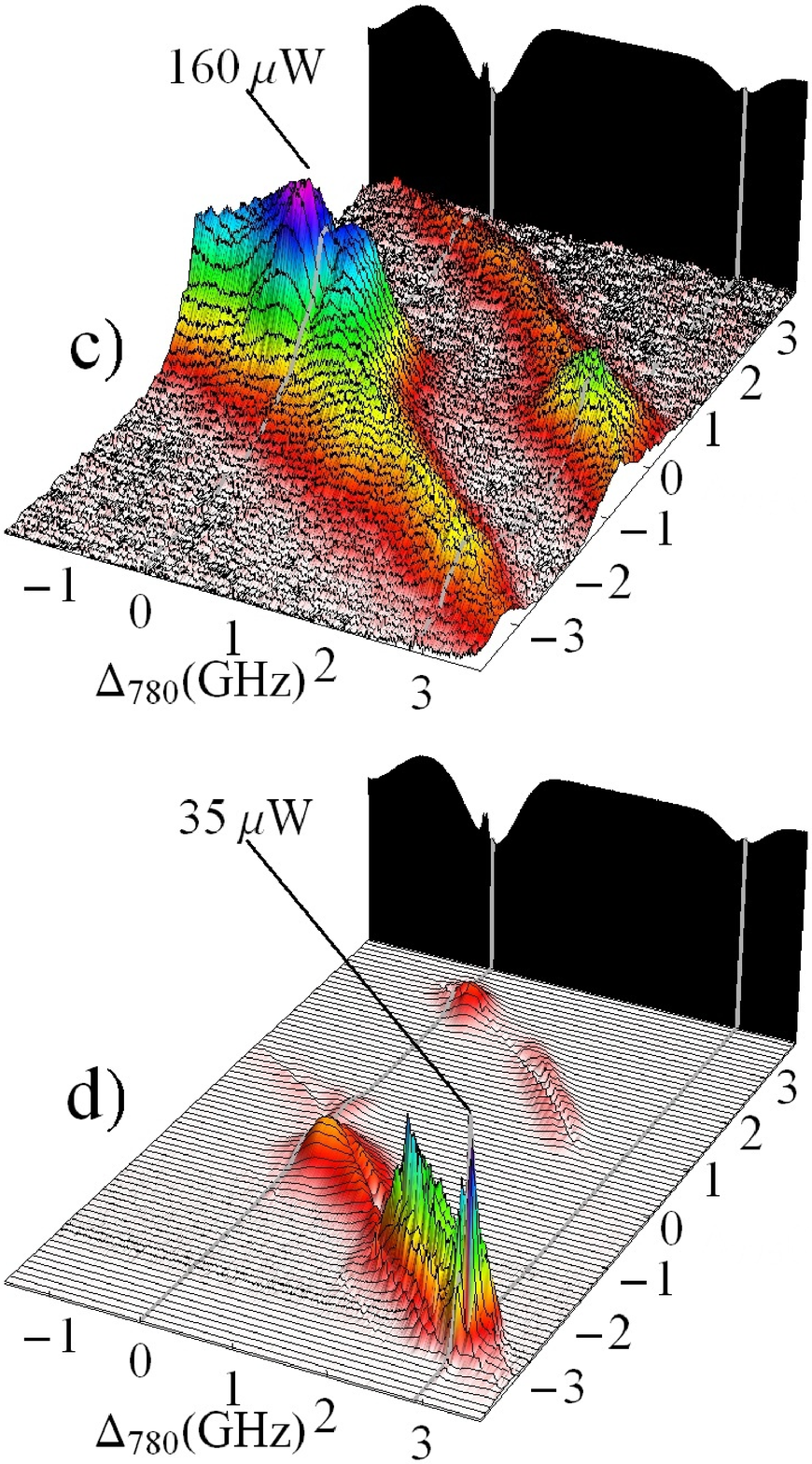}} \end{minipage}
\begin{minipage}{.352\columnwidth} \mbox{\includegraphics[clip,width=\columnwidth]{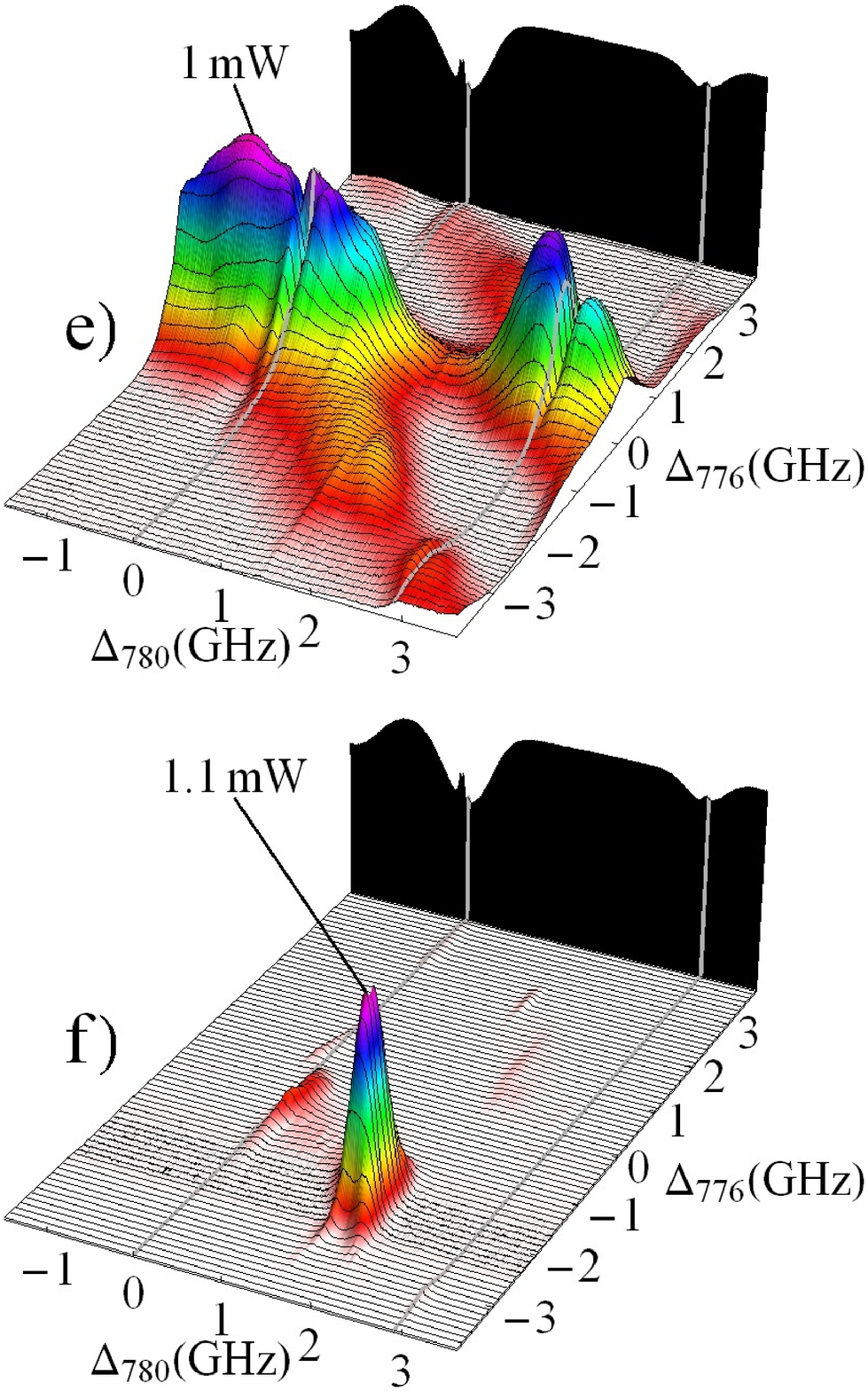}} \end{minipage}
\caption{Relative blue fluorescence (top row) and blue beam (bottom row) power as a function of 780 and $776\,$nm light detuning. Theoretical
simulations of the 5-level Bloch equations are shown in a) and b).  Graph a) shows the Doppler-broadened population in the 6P level corresponding to
blue fluorescence. Fig.~b) shows the Doppler-broadened coherences between ground and the 6P level, corresponding to coherent blue light generation.
The blue coherences are adjusted to account for Kerr lensing and absorption of the $780\,$nm pump beam. Graphs c) and d) show measurements at a cell
temperature of $90^{\circ}$C, and e) and f) at $120^{\circ}$C. At low pressure, coherent light is generated over a range of detunings at two-photon
resonance.  At high pressure Kerr-lensing of the pump beams and phase-matching become more pronounced, restricting blue light generation to a narrow
frequency window. The backdrops show $780\,$nm saturated absorption in a cold cell. \label{temps}}
\end{figure}

In order to investigate the up-conversion process we have measured fluorescent and coherent blue light generation over a wide range of input beam
frequencies, see Fig.~\ref{temps}c)-f).  Data was taken for low and high Rb temperatures of $90^{\circ}$C and $120^{\circ}$C respectively,
corresponding to vapor pressures of $1\times10^{-4}\,$mbar and $9\times 10^{-4}\,$mbar. The frequency of the $780\,$nm beam was measured via
saturated absorption spectroscopy in a room temperature Rb cell, and that of the $776\,$nm beam by monitoring the two-photon absorption of a weak
$776\,$nm probe beam counter-propagating through the hot cell (Fig.\ \ref{setup}a).

We have modeled fluorescence and coherent light generation (Fig.~\ref{temps}a) and b) from optical Bloch equations for the 5 atomic levels, using a
similar model to Ref.~\cite{scholten}, evaluated with Doppler broadening (FWHM$\sim580\,$MHz for the 776 and $780\,$nm light at $\sim 100^{\circ}$C)
and for Rabi frequencies ($\Omega_{780}=1.4\,$GHz, $\Omega_{776}=0.4\,$GHz, $\Omega_{IR}=30\,$MHz, $\Omega_{420}=20\,$MHz) similar to the estimated
experimental focal values in the low pressure cell. The model does not explicitly include propagation, and neglects the hyperfine structure of the
upper states. Nevertheless, numerical solutions for the steady state of the optical Bloch equations allowed us to interpret many of the observed
features. Fluorescence (top row of Fig.~\ref{temps}) arises from a transfer of population to the 5D level and subsequent fast decay to the 6P level.
Our model shows that this happens mainly when the system is pumped at two-photon resonance between the Stark-shifted atomic levels, and to a lesser
extent, for resonant driving of the $776\,$nm transition. The model (Fig.~\ref{temps}a) describes the observed fluorescence well (Figs.~\ref{temps}c
and e), even though it does not account for propagation effects or 780$\,$nm pump Kerr lensing which increase at higher temperature.

The generation of coherent blue light requires coherences between the ground state 5S hyperfine levels and the 6P level.
While the detuning of the blue (and IR) light is experimentally unknown, our simulations suggest that coherent blue light conversion occurs at three-photon resonance. For the (relatively) small blue light and IR intensities in the experiment 
the 6P level is unshifted, so three-photon resonance coincides with two-photon resonance. Phase-matching is a prerequisite of efficient coherent
light generation.  Our model shows that phase-matching $\Delta k=n_{780}k_{780}+n_{776}k_{776}-n_{IR}k_{IR}-n_{420}k_{420}\approx 0$ is satisfied
along most of the three-photon resonances. We model blue light generation (Fig. \ref{temps}b) as Doppler-broadened coherences adjusted to account for
Kerr lensing and absorption of the $780\,$nm pump beam. This is an approximation to a full analysis which would require the study of field and
density matrix propagation for non-collinear beams. At lower vapor pressure, we observe coherent blue light over a range of detunings near two-photon
resonance (Fig.~\ref{temps}d), reminiscent of observations in Ref.~\cite{scholten,hannaford}. Strikingly, when the Rb pressure in the cell is high,
the experiment shows an increase of blue beam generation by 1.5 orders of magnitude, however only over a restricted range of input detunings
(Fig.~\ref{temps}f), with highest conversion efficiency for two-photon (and three-photon) resonance at $\Delta_{780}=-\Delta_{776}=1.6\,$GHz. We
attribute the suppression of blue light generation outside this region to Kerr-lensing of the $780\,$nm input beam. Furthermore, absorption of
$780\,$nm light contributes to the absence of light generation from the $^{87}$Rb isotope.

\begin{figure}[!b]
 \begin{center}
 \mbox{\includegraphics[clip,width=.96\columnwidth]{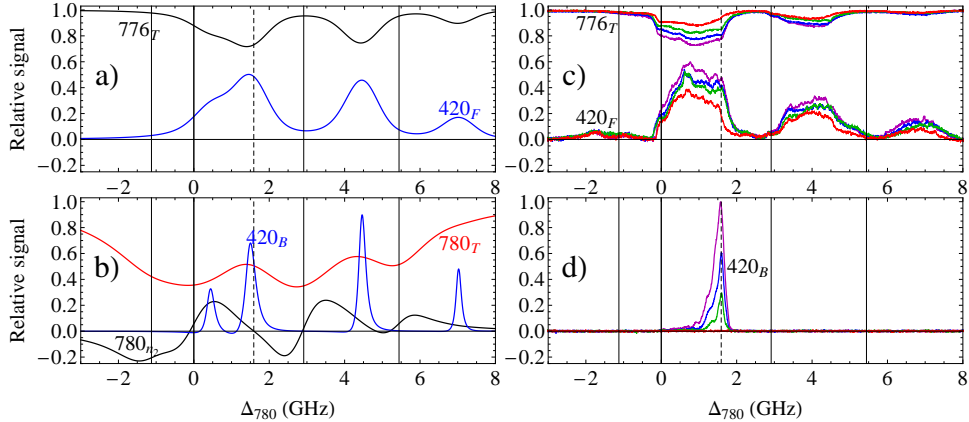}}
 \end{center}
\caption{Coherent and fluorescent blue light as a function of $\Delta_{780}$ for $\Delta_{776}=-1.6\,$GHz. Graph a) shows theoretical blue
fluorescence ($420_\mathrm{F}$, blue) as well as transmission of the $776\,$nm ($776_\mathrm{T}$, black) pump beam, including Doppler-broadening.
Graph b) shows the blue beam coherences ($420_\mathrm{B}$, blue), as well as the Doppler-broadened theoretical transmission ($780_\mathrm{T}$, red)
and Kerr-lensing ($780_{n_2}\propto \Delta n_{780}/\Delta I_{780}$ intensity-dependent refraction, black) of the $780\,$nm beam, which favors the
coherence at $1.6\,$GHz. Measurements of blue fluorescence accompanied by $776\,$nm absorption c), and blue beam generation d) are shown for
different input polarizations: counter-circular (red), crossed-linear (green), co-linear (blue) and co-circular (purple). Vertical lines indicate
$^{85}$Rb and $^{87}$Rb $780\,$nm resonances.} \label{780detuning}
\end{figure}

A scan (Fig.~\ref{780detuning}) of the $780\,$nm input frequency, keeping the $776\,$nm detuning constant at the optimized value of
$\Delta_{776}=-1.6\,$GHz, shows clearly the strong frequency dependence of fluorescence and blue light generation. The theoretical model
(Fig.~\ref{780detuning}a) of blue fluorescence accompanied by $776\,$nm absorption agrees well with the experiment (Fig.~\ref{780detuning}c).
Coherences on the 420$\,$nm transition between the 5S level and 6P level are present at two-photon resonance for the two hyperfine ground levels of
$^{85}$Rb and $^{87}$Rb, and mark out four frequency regions of potential blue light generation, Fig.~\ref{780detuning}b). Kerr-lensing depletes and
deflects the $780\,$nm pump laser at three of these resonances, so that only the coherence at $\Delta_{780}=1.6\,$GHz survives.  The same detuning
also exhibits a local minimum of absorption of the $780\,$nm input laser. We note that Kerr lensing will change beam direction and hence also modify
the phase-matching condition. Increasing the atomic density and hence the optical path length, makes propagation effects and phase-matching more
critical and adds further complexity as more 5$\,\mu$m and 420$\,$nm fields are generated. Experimentally, the strongest blue light generation was
observed near a minimum of lensing for the 780$\,$nm beam, Fig.~\ref{780detuning}d).

We finally report the crucial effect of input polarization on blue light generation, shown in Fig.~\ref{780detuning}c) and d).  For these experiments
the PBS in the dashed region of Fig.\ \ref{setup}a) had to be removed. We simultaneously monitored the blue beam power (PD$_{\rm B}$), the blue
fluorescence in the cell (PD$_{\rm F}$), absorption of the $776\,$nm beam (PD$_{776}$), and the detuning of the $780\,$nm laser (PD$_{\rm SAS}$).
Measurements were taken for four different relative input polarizations of the $776\,$nm and $780\,$nm input beams: co-linear, crossed-linear,
co-circular and counter-circular. Care was taken to compensate for grating polarization-dependence in the $776\,$nm pump beam measured at PD$_{776}$.
By analyzing the $776\,$nm laser absorption (Fig.~\ref{780detuning}c), we can estimate that it is possible to create a coherent $420\,$nm photon for
every six $776\,$nm photons absorbed. Note that the blue beam generation is suppressed by a factor of 500 when counter-circular polarization is used,
although there is still a similar amount of blue fluorescence and $776\,$nm beam absorption, indicating that the transfer of population to levels
5D$_{5/2}$ and 6P$_{3/2}$ is not forbidden by two- and three-photon absorption selection rules. This implies that the excitation of three-photon
coherences between the 5S$_{1/2}$ ground state and the 6P$_{3/2}$ state depend crucially on polarization. Optical pumping favors blue light
generation for the case of co-circularly polarized pump beams, however, the inhibition of blue light generation for counter-circular polarization
suggests an interference effect. This may be attributed to a cancelation of the coherences due to the phase of the atomic dipoles.
One can envisage that this property will be useful for applications in optical switching, where flipping $780\,$nm (or $776\,$nm)pump polarization
would switch the blue beam power.

As our system is not saturating with pump power, increasing the pump power or using a build-up cavity could generate tens of mW of blue light. By
accessing higher lying $D_{5/2}$ levels one should be able to generate similar powers at UV wavelengths by transitions via the corresponding
$P_{3/2}$ levels. It is interesting to consider the application of such light sources for laser cooling where the narrower linewidth on these blue/UV
transitions will lead to much lower Doppler temperatures \cite{hall}. In initial experiments we have observed $420\,$nm linewidths of less than
$8\,$MHz, at the resolution limit of our etalon. The scheme should generalize to all alkali metals, and has recently been realized in cesium
\cite{close}.

In conclusion, we have observed a polarization-dependent factor of 26 increase in blue light conversion efficiency under comparable conditions to
previous experiments. Our observations of the frequency-dependent characteristics of the $776\,$nm beam absorption, blue fluorescence and blue beam
power show good agreement with our theoretical framework, particularly for low experimental temperatures, indicating the importance of three-photon
resonance coinciding with vanishing susceptibility on the $780\,$nm transition. In the low Rb pressure regime with the weaker blue beam afforded by
crossed-linear polarizations we observe similar experimental behavior (frequency dependence and conversion efficiency) as observed in
Refs.~\cite{hollberg,scholten} and we attribute our increased efficiency to polarization enhancement as well as a higher apparent Rb vapor pressure.
It is clear that propagation must be included for proper modeling of the system at high temperatures due to the strong absorption/refraction
processes at work in the cell. There are still many open questions in this surprisingly rich atomic vapor system. We envisage future experiments in
isotopically enhanced vapor cells (to simplify the system's absorption-emission characteristics) and sapphire vapor cells which will allow us access
to the `missing link' of the closed loop structure \cite{sonja1,sonja2}, the $5.2\,\mu$m light.

Thanks to N.\ Paterson, S.\ Clark and N.\ Houston for experimental contributions. We gratefully acknowledge discussions with P.\ Siddons and I.~G.\
Hughes. SFA is a RCUK Research Fellow.

\end{document}